\documentclass[a4paper,11pt]{article}
\usepackage[utf8]{inputenc}
\usepackage[T1]{fontenc}
\usepackage{lmodern}
\usepackage{setspace}
\usepackage[super,comma]{natbib}
\usepackage{authblk}
\usepackage[a4paper,margin=3.5cm]{geometry}
\usepackage{graphicx,verbatim,amsmath,amssymb}

\usepackage{color}
\definecolor{theblue}{rgb}{0.086,0.173,0.514}

\usepackage[bookmarks,bookmarksopen,pdfdisplaydoctitle,colorlinks,
  urlcolor=theblue,linkcolor=theblue,citecolor=theblue,
  hyperindex=false,hyperfootnotes=false,
  pdftitle={},pdfauthor={},pdfsubject={},pdfkeywords={}
]{hyperref}

\DeclareRobustCommand\mailto[1]{\href{mailto:#1}{\nolinkurl{#1}}}

\definecolor{thered}{rgb}{0.8,0,0}

\begin{document}

\title{Spectral purity transfer between optical\\ 
  wavelengths at the $10^{-18}$ level}

\author{Daniele Nicolodi}
\author[1]{B\'ereng\`ere Argence\thanks{Present address: Laboratoire de
    Physique des Lasers, Université Paris 13, Sorbonne Paris Cité,
    CNRS, 99 Avenue Jean-Baptiste Clément, 93430 Villetaneuse, France}}
\author[1]{Wei Zhang\thanks{Present address: JILA, National Insitiute of
    Standard and Technology and University of Colorado, Boulder,
    Colorado, 80309-0440 USA}}
\author[1]{Rodolphe Le Targat}
\author[1,2]{Giorgio Santarelli}
\author[1]{Yann Le Coq\thanks{Corresponding author: \mailto{yann.lecoq@obspm.fr}}}
\affil[1]{%
  LNE-SYRTE, Observatoire de Paris, CNRS, UPMC,
  61 Avenue de l'Observatoire, 75014, Paris, France}
\affil[2]{%
  Laboratoire Photonique, Numérique et Nanosciences, UMR
  5298 Université de Bordeaux 1, Institut d'Optique and CNRS, 351
  Cours de la Libération, 33405 Talence, France}

\maketitle

\doublespacing

Ultra-stable lasers and optical frequency combs have been the enabling
technologies for the tremendous progress of precise optical
spectroscopy in the last ten years~\cite{hall-revmodphys-2006,
  hansch-revmodphys-2006}. To improve laser frequency stabilization
beyond the thermal-noise fundamental limit of traditional
room-temperature high-finesse optical
cavities~\cite{bergquist-prl-1999}, new solutions have been recently
developed~\cite{ludlow-natphot-2011, thorpe-natphot-2011,
  kessler-natphot-2012, cole-natphot-2013}. These being complex and
often wavelength specific, the capability to transfer their spectral
purity to any optical wavelengths is highly desirable.  Here we
present an optical frequency comb based scheme transferring a $4.5
\times 10^{-16}$ fractional frequency stability from a 1062~nm
wavelength laser to a 1542~nm laser. We demonstrate that this scheme
does not hinder the transfer down to $3 \times 10^{-18}$ at one
second, two orders of magnitude below previously reported work with
comparable systems~\cite{coddington-natphot-2007,
  nakajima-optexpr-18-1667, hagemann-ieee-99-1,
  inaba-optexpr-21-7891}. This exceeds by more than one order of
magnitude the stability of any optical oscillator demonstrated to
date~\cite{kessler-natphot-2012}, and satisfies the stability
requirement for quantum-projection-noise limited optical lattice
clocks~\cite{ye-prl-2012}.

\clearpage

Optical frequency combs provide a phase coherent link across the
optical and microwave frequency domains. For instance they were used
for the generation of ultra-low phase noise microwave
signals~\cite{fortier-natphot-2011, lecoq-apl-96-211105,
  lecoq-optlett-36-3654}, that allow the interrogation of atomic
fountain clocks at the quantum-projection-noise
limit~\cite{millo-apl-94-141105, weyers-pra-79-031803}.  Here we
present a solution exploiting an optical frequency comb (OFC) based on
an Erbium-doped-fiber femtosecond oscillator centered at 1560~nm for
the phase locking of two optical oscillators. We demonstrate
spectral purity transfer without frequency stability degradation from
a $1062$~nm master laser stabilized by an high-finesse optical cavity,
independently characterized at the $4.5 \times 10^{-16}$ stability
level~\cite{millo-pra-79-053829}, to a $1542$~nm slave laser.

Bridging the large wavelength difference between the slave and the
master lasers require spectral broadening of the femtosecond laser
output to obtain a sufficiently wide frequency comb.  In fiber-based
optical frequency comb systems, usually, the femtosecond oscillator
output is amplified and spectrally broadened in dedicated
branches to obtain a phase coherent OFC output centered in the
spectral region of each cw laser.  The relative phase between the cw
lasers is then obtained by beating each of them to the output of the
dedicated branch. The amplification and spectral broadening of the comb's
output in dedicated branches introduces differential phase noise that
degrades the phase comparison and thus the frequency stability
transfer.  Typically, differential phase noise prevents frequency
comparison of the master and slave lasers with a fractional resolution
better than about $1 \times 10^{-16}$ near one second integration
time\cite{nakajima-optexpr-18-1667, hagemann-ieee-99-1,
  inaba-optexpr-21-7891}.

\begin{figure}
\includegraphics[width=\textwidth]{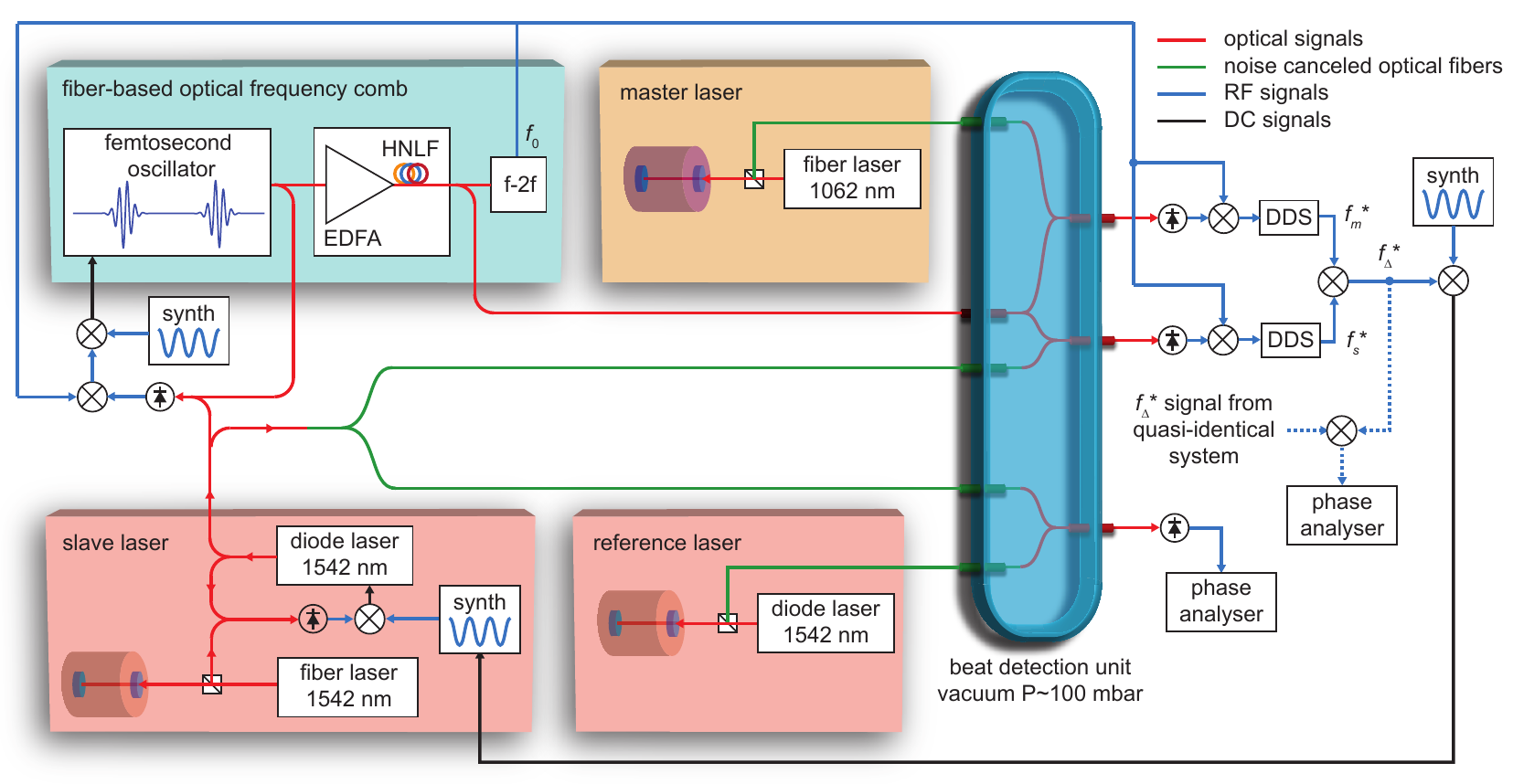}
\caption{Experimental setup. EDFA: Erbium-doped fiber amplifier; HNLF:
  highly non-linear fiber; Synth: radio-frequency synthesizer; DDS:
  direct digital synthesizer. This setup is used for transferring the
  spectral purity from the master laser at $1062$~nm wavelength to the
  slave laser at $1542$~nm wavelength with a heterodyne phase-locking
  technique. To demonstrate the stability improvement, the resulting
  oscillator is characterized against the $1542$~nm reference laser
  with a phase noise analyzer. To characterize the limit of the
  system, we compare two such comb-based systems and compare their
  results with the phase noise analyzer, with a setup illustrated by
  the dashed lines.}
\label{f:scheme}
\end{figure}

To overcome this limitation we beat both cw laser with a single
spectrally broadened comb's output obtained from the f-2f
interferometer unit used for the detection of the carrier-envelop
offset frequency $f_0$.  Spreading the available optical power over a
broad spectrum results in low intensity of the comb teeth, and thus
the optical beatnotes with each cw laser have relatively low
signal-to-noise ratio (SNR).  Because of the low SNR, narrow bandwidth
detection of the beatnotes is required. This is made possible
operating the comb in the ``narrow-linewidth regime'' in which each
tooth has nearly the same spectral purity as the cw laser to which the
comb is locked~\cite{zhang-ieee-2012}.  The experimental setup is
illustrated in figure~\ref{f:scheme}.

The comb's repetition rate $f_{rep}$ is phase locked directly beating
a fraction of the non amplified femtosecond oscillator output with the
1542~nm cw laser.  The high SNR of the resulting optical beatnote and
a fast intra-cavity electro-optic modulator (EOM) actuator allow a
phase-lock loop with a bandwidth exceeding~$1$~MHz, thus permitting to
operate the comb in the narrow-linewidth regime.  In the case where
neither the master or the slave laser have a wavelength contained in
the spectrum of the femtosecond oscillator output, an auxiliary stable
cw laser oscillator is required to lock the comb's repetition rate.
Another fraction of the femtosecond oscillator output is amplified in
a Erbium-doped fiber amplifier (EDFA) and spectrally broadened by a
highly non linear fiber (HNLF) to obtain an octave spanning spectrum
from 1~$\mu$m to 2~$\mu$m.  The comb's EDFA-HNLF output is separated
with a fibered dichroic splitter in the spectral components near
1062~nm and 1542~nm, which are combined with the light stemming from
the master and slave lasers respectively, transported via actively
noise canceled fibers to a bead detection unit. The resulting
beatnotes are photo-detected and mixed with the carrier-envelope
offset frequency to obtain the $f_0$-free signals $f_m = \nu_m - N_1
f_{rep}$ and $f_s = \nu_s - N_2 f_{rep}$, related respectively to the
instantaneous frequency of the master laser $\nu_m$ and of the slave
lasers $\nu_s$. The $f_s$ and $f_m$ signals are filtered through
tracking oscillators with a bandwidth of roughly $5$~kHz and used to
clock two direct digital synthesizers implementing frequency division,
resulting in $f_m^* = f_m / M_1$ and $f_s^* = f_s / M_2$ signals. The
$M_1$ and $M_2$ divisors are chosen such as $M_1/M_2 = N_1/N_2$. This
allows the generation of a $f_\Delta^* = f_m^* - f_s^*$ signal
independent of the repetition rate $f_{rep}$ and therefore immune to
its phase fluctuations, as in the ``transfer oscillator
technique''~\cite{telle-appphysb-2002}. Because $N_1/N_2 \simeq
\nu_m/\nu_s$, the signal at frequency $f_\Delta^*$ implements properly
the heterodyne phase comparison between the 1542~nm and 1062~nm
wavelength lasers. By comparing $f_\Delta^*$ with a fixed frequency
reference, we derive a phase error which is used to correct the
frequency of the slave laser, thereby implementing a phase-locked loop
of the 1542~nm slave laser on the 1062~nm master laser.  The
phase-locked loop bandwidth is limited to roughly $5$~kHz by the
bandwidth of the tracking oscillators, and requires pre-stabilization
of the slave laser to a linewidth narrower than roughly $100$~Hz.

\begin{figure}
\centering
\includegraphics[width=\textwidth]{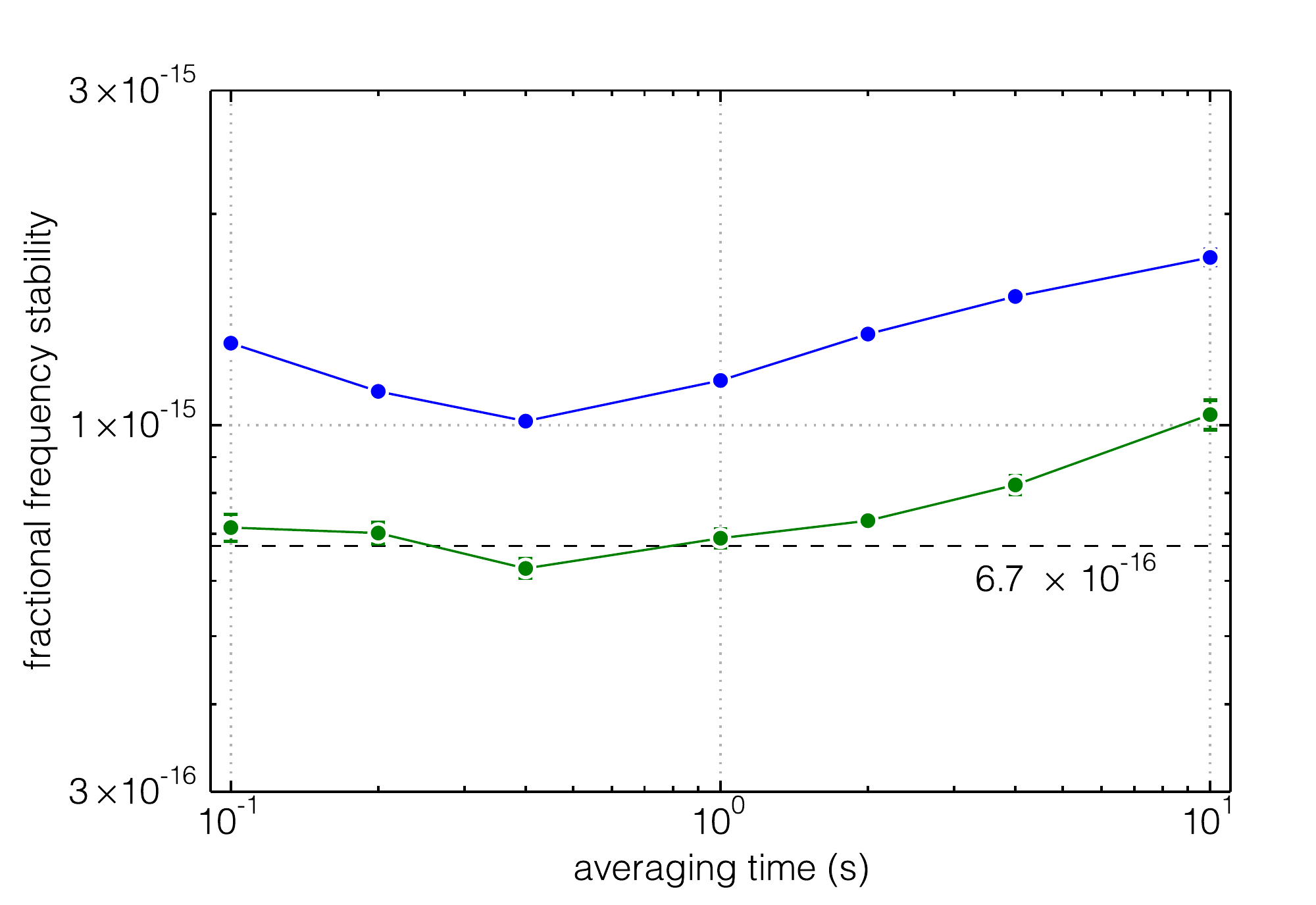}
\caption{Relative frequency stability of the beatnote between the
  $1542$~nm slave laser and the $1542$~nm reference laser (standard
  Allan deviation, $5$~Hz measurement bandwidth). Blue: the slave
  laser is pre-stabilized only, no feed-back via the comb is
  applied. Green: the slave laser is phase-locked to the master laser
  through the comb, the slave laser copies the performance of the
  master laser. Error bars, when non visible, are smaller than the
  graphical size of the marker. The apparent stability degradation at
  longer integration times is due to residual frequency drift between
  the slave and reference lasers.}
\label{f:adev}
\end{figure}

To demonstrate stability and spectral purity improvement arising from
the slave-to-master phase lock loop, we beat the slave laser to an
independent cavity stabilized $1542$~nm reference laser, independently
characterized at the $5.0 \times 10^{-16}$ stability
level~\cite{argence-optexpress-2012}. The fractional frequency
stability of the resulting beatnote is presented in
figure~\ref{f:adev}. It is as low as $6.7 \times 10^{-16}$ for
integration times between $0.1$~s and $1.0$~s. In the assumption of
uncorrelated frequency noise, this demonstrates the transfer of
spectral purity from the 1062~nm master laser to the 1542~nm slave
laser with negligible additional frequency instability.

\begin{figure}
\centering
\includegraphics[width=\textwidth]{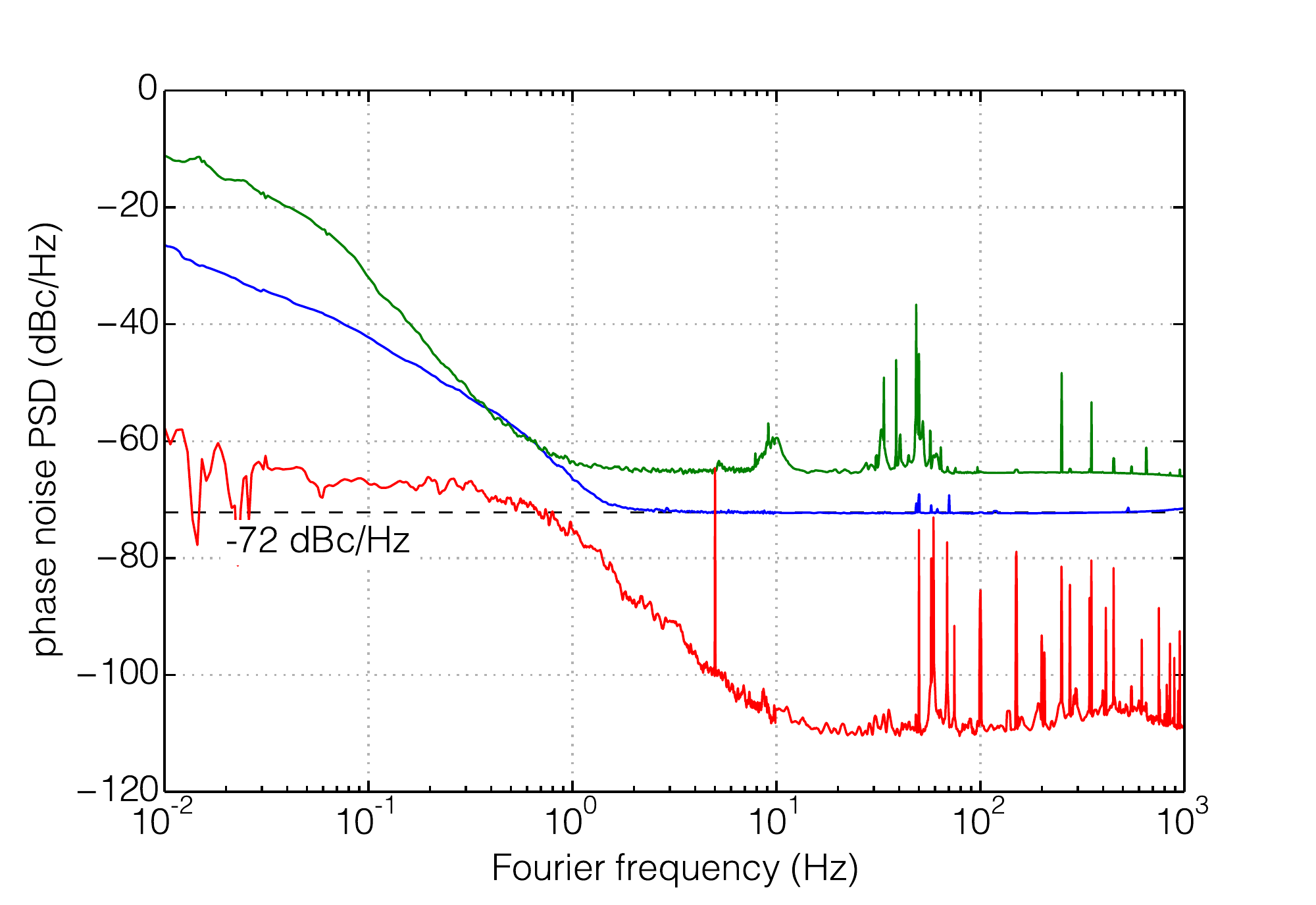}
\caption{Blue: phase noise single-sided power spectral
  density (PSD) limit of the spectral purity transfer setup
  assessed comparing the phase difference between the same
  master and slave laser measured by two independent
  quasi-identical setups, assuming that each system contributes equal
  and uncorrelated phase noise. The $-72$~dBc~Hz$^{-1}$ phase
  white noise plateau is defined by the SNR of the optical
  beatnotes which is limited by the low available comb's light power
  in the cw lasers spectral regions. Green: phase noise of
  the spectral purity transfer setup assessed measuring the phase
  difference between two slave lasers phase-locked to the same
  master laser through two independent setups, assuming that each
  system contributes equal and uncorrelated phase noise.  The
  increase of the white noise plateau to $-65$~dBc~Hz$^{-1}$ stems
  from decreased SNR of the optical beatnotes due to technical
  reasons.  The low frequency noise increase is due to phase-noise
  contribution from the interferometer used to beat the two slave
  lasers, which includes a few meters of non noise canceled optical
  fibers. Red: typical residual phase noise of the slave lasers
  phase-locking (the $5$~Hz spectral line is a calibration tone).}
\label{f:phase-noise-psd}
\end{figure}

\begin{figure}
\centering
\includegraphics[width=\textwidth]{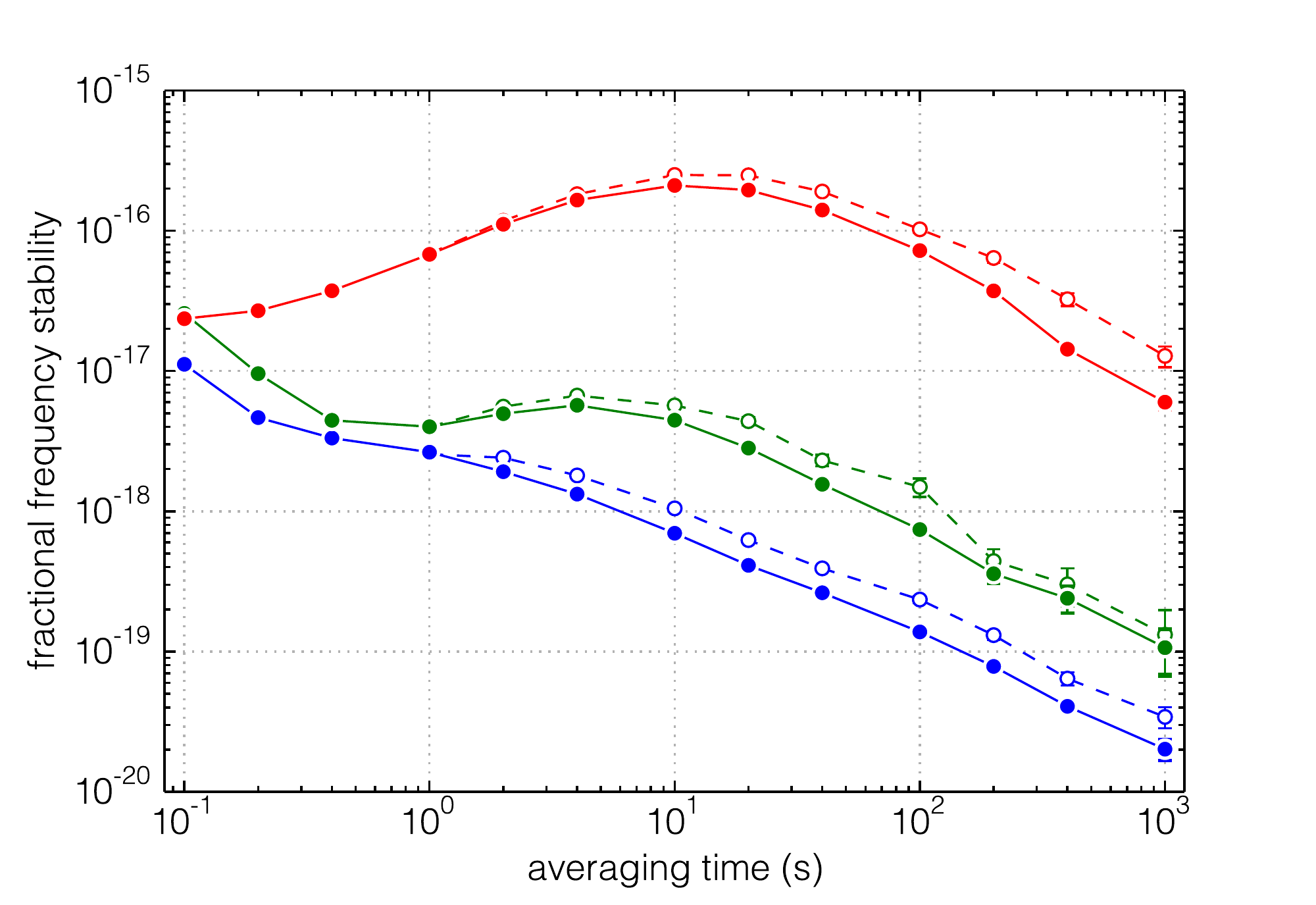}
\caption{Frequency stability limit of the setup (Allan
  deviation, open markers, $0.5$~Hz measurement bandwidth; modified
  Allan deviation, filled markers). Blue: Frequency measurement
  capability assesment configuration.  The frequency stability is
  determined by comparing the phase difference between the same
  master and slave laser measured by two independent quasi-identical
  setups, assuming equal and uncorrelated contribution from each.
  Green: Full spectral purity transfer configuration. The frequency
  stability is assessed measuring the phase difference between two
  slave lasers locked to the same master laser through two
  independent setups, same assumption as previously.  The
  degradation of the frequency stability from the blue curve is due,
  at short averaging times, mostly to a technical reduction of the
  optical beatnotes SNR. At long averaging times, it is mostly
  attributed to the contribution from the interferometer used to
  beat the two slave lasers, which includes a few meters of non
  noise canceled optical fibers.  Red: configuration representative
  of a multi-branch setup. Error bars, when non visible, are smaller
  than the graphical size of the marker.}
\label{f:mdev}
\end{figure}

To evaluate the optical frequency measurement capability of
our system, we compare the $f_\Delta^*$ signals obtained from two
independent quasi-identical setups measuring the phase
difference between the same master and slave laser. The two setups
differ only in minute details of the comb's repetition rate phase-lock
loop.  From the phase difference between the two $f_\Delta^*$ signals,
assuming that each system contributes equal and uncorrelated phase
noise, we assess the phase noise added by the OFC setup in the
frequency stability transfer process.  Figure~\ref{f:phase-noise-psd}
and~\ref{f:mdev} present the phase noise power spectral density
and the corresponding fractional frequency stability.

The measured additive phase noise exhibits white noise plateau at
$-72$~dBc~Hz$^{-1}$ that extends down to $2$~Hz Fourier
frequency. This white phase noise level is defined by the SNR of the
$f_m$ and $f_s$ beatnotes which is limited by the low available comb's
light power in the $1062$~nm and $1542$~nm spectral regions. Possible
improvements of the phase noise in this Fourier frequency region may
involve obtaining an higher SNR with amplification of the relevant
regions of the EDFA-HNLF output spectrum using semiconductors
amplifiers~\cite{fortier-prl-2006}, or employing sub-shot-noise gated
detection scheme~\cite{deschenes-pra-2013}.

On longer time scales, the system exhibits excess noise that limits
the fractional frequency stability to $3 \times 10^{-18}$ at one
second. This lower limit is close to one order of magnitude below any
previously reported work \cite{ma-ieeejqe-2007} and two order of
magnitude lower than fiber-comb based system
reports~\cite{coddington-natphot-2007, nakajima-optexpr-18-1667,
  hagemann-ieee-99-1, inaba-optexpr-21-7891}. The $2 \times 10^{-20}$
stability at $1000$~s is also the lowest reported long term stability
on OFC systems~\cite{ma-ieeejqe-2007}. Possible residual
sources of the low frequency excess noise are incomplete noise
suppression of the optical fiber links, and laser power
fluctuations.  Additionally, despite the beatnote detection is
performed in an carefully designed vacuum setup, providing good
isolation from acoustic and thermal disturbances, we cannot completely
rule out those as relevant sources of phase noise. The detection of
$f_m$, $f_s$, and $f_0$ from the same spectrally broadened comb output
is effective in suppressing the phase fluctuations introduced in the
EDFA and HNLF when those are wavelength independent or scale linearly
with optical frequency.  Higher order terms are however not canceled
and may contribute to the measured excess phase noise.

To fully characterize the optical frequency stability
transfer, we use the two independent setups to phase-lock two slave
lasers to the same master laser. We assess the phase noise of the
transfer process from that of the beatnote between the two slave
lasers, assuming equal and uncorrelated contribution from each
system. Figure~3 and~4 present the phase noise power spectral
density and the corresponding fractional frequency stability.

For technical reasons, in this configuration, the SNR of the
$f_m$ and $f_s$ beatnotes is degraded, which increases the white
noise plateau to $-65$~dBc~Hz$^{-1}$ and limits the fractional
frequency stability to $4 \times 10^{-18}$ at one second. Long term
stability is affected by the noise contribution of the
interferometer beating the two slave lasers which includes a few
meters of non noise canceled fibers. We still demonstrate fractional
frequency stability of $1\times 10^{-19}$ at $1000$~s.

Those results are obtained with a reliable fiber-based system,
compatible with long-term continuous and autonomous operation, that
make it suitable for the use in metrology experiments.  We routinely
operated the system in a research laboratory environment to obtain
unsupervised, continuous, phase-slip free operation limited only by
anthropogenic disturbances.  No fundamental limitations preventing the
system to be operated continuously have been identified.

The presented technique is readily applicable for phase-locking any
laser oscillator in the spectral region between 1~$\mu$m and 2~$\mu$m
where the comb's EDFA-HNLF output provides non-negligible optical
power. It and can be extended to the visible spectral region via
second harmonic generation of the comb's output. Moreover,
ultra-stable lasers in the visible spectral regions used for metrology
applications are often obtained via second harmonic generation.  The
proposed scheme is therefore directly applicable to most of the
existing experiments.

Identifying unequivocally the dominant sources of low frequency noise
and developing strategies to circumvent them will require further
work. Nonetheless, the current $3 \times 10^{-18}$ level exceeds by
more than one order of magnitude the performance of any oscillator
demonstrated to date~\cite{ludlow-natphot-2011, thorpe-natphot-2011,
  kessler-natphot-2012, cole-natphot-2013} and is compatible with the
requirement of quantum-projection-noise limited optical lattice atomic
clocks~\cite{ye-prl-2012}. Transferring frequency stability from an
optical oscillator at arbitrary wavelength to a laser with wavelength
centered in the femtosecond oscillator output spectrum, and thus
compatible with wide bandwidth phase-locking of the comb's repetition
rate, the proposed technique can also benefit OFC based low
phase-noise microwave signal generation, which is limited, close to
the carrier, by the phase noise of the reference optical
oscillator.~\cite{fortier-natphot-2011}.

By demonstrating the capability to transfer their spectral purity to
any frequency comb accessible wavelengths, this work strengthens the
quest for extremely-high-stability lasers at implementation-specific
wavelengths.

\section*{Methods}
\label{s:methods}

\paragraph{Laser systems.}

The master laser is an ultra-stable 1062.5~nm Yb-doped fiber-laser
locked via the Pound-Drever-Hall (PDH) scheme to an ultra-stable
Fabry-Perot cavity, obtaining fractional frequency stability~$4.5
\times 10^{-16}$ at one second~\cite{millo-pra-79-053829}. The slave
laser is a 1542.5~nm diode laser pre-stabilised by offset
phase-locking to an Er-doped fiber-laser which is itself stabilised by
PDH locking to a Fabry-Perot cavity. To demonstrate our setup
capability, we voluntarily degraded to roughly $1 \times 10^{-15}$ at
one second the fractional frequency stability of the slave laser
transmitting it through a section of non noise-canceled optical
fiber.  We stress that such level of pre-stabilization is not
required: the phase-locked loop locking the slave to the master
laser has a bandwidth of roughly $5$~kHz, and therefore
pre-stabilization of the slave laser to a linewidth narrower than
roughly $100$~Hz is sufficient. This requirement can be
easily met by locking the slave laser to a simple optical cavity or to
an optical fiber-delay line~\cite{kefelian-optlet-2009}, obtaining a
significant reduction of the complexity of the setup.  The reference
laser is a 1542.5~nm Er-doped fiber laser locked via the PDH scheme to
an ultra-stable Fabry-Perot cavity, obtaining fractional frequency
stability~$5.0 \times 10^{-16}$ at one
second~\cite{argence-optexpress-2012}.

\paragraph{Fiber-based optical frequency comb.}

The optical frequency comb is a commercial core fiber based optical
frequency comb based on a femtosecond Erbium-doped fiber laser
equipped with an intra-cavity electro-optic modulator (EOM). The EOM
provides feedback on the comb's repetition rate $f_{rep}$ with a
bandwidth of roughly $1$~MHz and allows to lock the comb to the slave
laser in the narrow-linewidth regime in which each tooth has nearly the
same spectral purity as the ultra-stable cw laser. The repetition rate
phase lock technique is described in reference~\cite{zhang-ieee-2012}.
Briefly, the femtosecond laser output is filtered around $1542$~nm
with a bandwidth of $0.8$~nm in an optical add-drop multiplexer (OADM)
and beated with the cw slave laser of optical frequency $\nu_{cw}$ leading
to a beat note $f_b = \nu_{cw} - N f_{rep} - f_0$, where $N$ is an
integer addressing the comb's tooth closer to the optical frequency
$\nu_{cw}$. A built-in f-2f interferometer unit produces the
carrier-envelope offset frequency $f_0$ which is mixed with $f_b$ to
produce a $f_0$-independent signal of frequency $\nu_{cw} - N
f_{rep}$. This signal is frequency divided by eight and mixed with a
fixed frequency from a synthesizer to obtain a phase error
signal. This error signal is processed in an analog loop filter,
implementing proportional and multiple integrator gains, and used to
correct the comb's repetition rate by acting on the intra-cavity EOM
and on a piezo-electric actuator controlling the cavity length. Both $f_b$ and
$f_0$ signals exhibit SNR greater than $40$~dB in $1$~MHz
bandwidth. The large SNR allows to use the whole feedback bandwidth
capability of the EOM actuator, such that the comb reaches the
narrow-linewidth regime.

\paragraph{Transfer oscillator technique combined with narrow-linewidth regime.}

Due to the limited light power available at $1062$~nm and
$1542$~nm in the EDFA-HNLF comb's output, the $f_m$ and $f_s$
signals present SNR smaller than $10$~dB in $1$~MHz bandwidth, limited
by the white noise floor of the detection chain. The small SNR makes
those signals not suitable to lock the comb in the narrow-linewidth
regime: SNR greater than $30$~dB in the lock bandwidth is normally
required for proper phase-lock loop operation.  Nonetheless, because
the comb is independently set in the narrow-linewidth regime, as
described in the previous section, we are able to track the low SNR
beatnotes with $5$~kHz of bandwidth, thereby generating clean
radio-frequency signals compatible with the high quality digital
signal processing that lies at the heart of the transfer oscillator
technique~\cite{telle-appphysb-2002}.

Even though the comb's repetition rate $f_{rep}$ is tightly phase
locked to the $\nu_{cw}$ $1542$~nm slave laser, it may exhibit
residual noise due to an imperfect and bandwidth limited servo
loop. The carrier-envelope frequency offset $f_0$ is not locked and,
in a laboratory well-controlled environment, it may evolve in a few
MHz range.  The transfer oscillator technique effectively suppresses
the effect of both frequency fluctuations on the master and slave
laser phase comparison.  We note however that spectral phase evolution
produced in the femtosecond laser scaling with optical frequency at
order higher than linear is not canceled.

We observe that, if neither the master or the slave laser happen to
have a wavelength contained in the spectrum of the femtosecond
oscillator output (approximately 30 nm wide around 1560 nm, where it
is possible to obtain a beatnote with large signal-to-noise ratio,
compatible with wide phase-locking bandwidth of the comb), an
auxiliary stable cw laser oscillator is required to lock the optical
frequency comb in the narrow-linewidth regime.  Due to the bandwidth
limitations exposed above, the combs teeth should not be wider than
about $100$~Hz, to do not impact on the performance of the system. This
can be realized via tight locking of the repetition rate to an
auxiliary cw laser with the same linewidth, which can be obtained with
laser stabilization on fiber spool delay
lines~\cite{kefelian-optlet-2009} significantly reducing the
complexity of the setup.

\paragraph{Phase noise added by the EDFA-HNLF system.}

The optical phase noise introduced in the EDFA-HNLF system is
wavelength dependent.  For example, optical path length fluctuations
introduce phase noise linear in the optical frequency, while optical
amplifier gain fluctuations and refraction index changes may introduce
phase noise with more complex dependency on the wavelength. 

Measuring the comb-cw laser beatnotes $f_m$ and $f_s$, and the
carrier-envelope frequency offset $f_0$ in the same EDFA-HNLF branch,
phase noise components with up to linear dependency on the optical
frequency cancel out when the $f_m$ and $f_s$ signals are combined to
generate the $f_\Delta^*$ signal. We note however that higher order
spectral phase evolution produced in the EDFA-HNLF are not canceled.

The effect of the transfer oscillator technique applied to the $f_m$,
$f_s$ and $f_0$ beatnotes measured from a single EDFA-HNLF branch is
therefore twofold: it removes the effect of residual $f_{rep}$ and
$f_0$ fluctuations, and provides a comparison insensitive to the
EDFA-HNLF added spectral phase noise up to linear order in optical
frequency.

\paragraph{Beatnote detection.}

The beatnote detection is realized in independent, carefully
designed, all fibered, beat detection units, enclosed in separate
vacuum chambers (one for each optical frequency comb setup),
providing thermal and vibration passive isolation.  Light from the
slave laser, the master laser, and the comb's EDFA-HNLF output are
sent to the beat detection units via single mode optical fibers and
manual polarization controllers used to match light polarization and
maximize the interferometers contrast. Optical fibers transmitting the
light from the cw laser sources are equipped with fiber-noise
cancelers~\cite{ma-optlett-19-1777} exhibiting feedback bandwidth
greater than $100$~kHz. Great care has been taken to keep to minimum
the length of the fibers which are either non-common path for the
$1542$~nm and $1062$~nm spectral components or not noise-canceled. The
beatnote detection unit is evacuated to about 100~mbar for acoustic
noise and thermal fluctuation isolation. No improvement has been
observed by further lowering the residual pressure.

\clearpage

\clearpage

\section*{Acknowledgments}

This work is partly funded by the Ville de Paris (Emergence(s) 2012
program) and CNES. D.N. and W.Z. gracefully acknowledge funding from
the EMRP IND14 project, which is jointly funded by the EMRP
participating countries within EURAMET and the European Union.

\section*{Author contributions}
All authors contributed extensively to the work presented in this paper.

\end{document}